\documentclass[journal]{IEEEtran}
\usepackage[a4paper, tmargin=1in, bmargin=1in, lmargin=0.8in,rmargin=0.8in]{geometry}
\usepackage[utf8]{inputenc}
\usepackage{graphicx}
\usepackage{mathtools}
\usepackage{pdflscape}
\usepackage{listings}
\usepackage{hyperref}
\usepackage{caption}
\usepackage{subcaption}
\usepackage{amsfonts}
\usepackage{amsmath}
\usepackage{gensymb}
\def\BibTeX{{\rm B\kern-.05em{\sc i\kern-.025em b}\kern-.08em
    T\kern-.1667em\lower.7ex\hbox{E}\kern-.125emX}}

\begin{document}

\title{A Learnable Distortion Correction Module for Modulation Recognition} 
\author{Kumar Yashashwi, Amit Sethi, Prasanna Chaporkar \\ Department of Electrical Engineering, \\ Indian Institute of Technology Bombay, India \\  \{kryashashwi, asethi, chaporkar\}@ee.iitb.ac.in}


\maketitle

\begin{abstract}
Modulation recognition is a challenging task while performing spectrum sensing in a cognitive radio setup. Recently, the use of deep convolutional neural networks (CNNs) has shown to achieve state-of-the-art accuracy for modulation recognition \cite{survey}. However, a wireless channel distorts the signal and CNNs are not explicitly designed to undo these artifacts. To improve the performance of CNN-based recognition schemes we propose a signal distortion correction module (CM) and show that this CM+CNN scheme achieves accuracy better than the existing schemes. The proposed CM is also based on a neural network that estimates the random carrier frequency and phase offset introduced by the channel and feeds it to a part that undoes this distortion right before CNN-based modulation recognition. Its output is differentiable with respect to its weights, which allows it to be trained end-to-end with the modulation recognition CNN based on the received signal. For supervision, only the modulation scheme label is used and the knowledge of true frequency or phase offset is not required.
\end{abstract}


\section{Introduction} \label{intro}
With an increasing number of users in wireless networks there has been an increasingly high congestion in the available spectrum making it a scarce asset. However, at times parts of the spectrum remain underutilized \cite{spectrum_policy}. This gives rise to the need for algorithms that can dynamically share the available spectrum. In the scenario of cognitive radio, spectrum sharing allows cognitive radio users (secondary) to share the spectrum bands of the licensed-band users (primary).
A key aspect of spectrum sharing is spectrum sensing \cite{spectrum_sensing}. Spectrum sharing involves white space detection based on which the secondary users (SU) communicate. Since the primary users (PU) opportunistically allow the secondary users to operate in an inactive frequency band originally allocated to the PUs, minimum time delay is desired \cite{timede}. Recent research efforts have been towards designing high-quality spectrum-sensing devices and algorithms to characterize the radio frequency (RF) environment, particularly for recognition of the modulation scheme. Distortion of the received signal due to channel fading effects makes the modulation recognition a challenging task. Hence, an algorithm that models and corrects the distortion caused by the channel should improve modulation recognition.
  
In the past few years, deep learning techniques have achieved state-of-the-art performance in pattern recognition tasks \cite{DL}. For the purpose of spectrum sensing different deep learning algorithms such as multilayer perceptron (MLP) \cite{nandi} and convolutional neural network (CNN) \cite{survey} have been proposed to recognize the modulation scheme from the given signal. In this work we introduce a module in a neural network to account for the random carrier frequency offset (CFO) and phase noise. Carrier frequency offset and phase noise are added randomly to the transmitted signal by the channel and as a result, the recognition accuracy reduces. For example, if frequency deviation in the transmittor is $10$ ppm above the centre frequency and the same is for the receiver, a CFO of $20$ ppm is induced effectively in the received baseband signal. If the carrier frequency is $4$ GHz, the CFO is up to $\pm 80$ kHz. Moreover, Doppler effect would further degrade the frequency offset if either the transmitter or reciever is moving. To tackle such problems, we propose a correction module (CM) to undo the effect of random frequency and phase noise without any prior information about these factors. To be more precise, the correction of CFO and phase noise is unsupervised. This idea is inspired from spatial transformer networks used in image recognition \cite{stn}. The CM when used with CNN improves the recognition accuracy for both high and low values of signal-to-noise ratio (SNR). We call this scheme CM+CNN.

The rest of the paper is organized as follows: Section \ref{background} discusses the related work and dataset generation method. In Section \ref{technique} we introduce our methodology. Section \ref{results} discusses our results and conclusions are in Section \ref{concl}.

\section{Background and related work}
\vspace{-0.04cm}
\label{background}
Techniques for determining modulation scheme have been depending on increasingly complex machine learning methods. Early work by Nandi \cite{nandi} implemented a decision theoretic and MLP approach for modulation recognition. A hierarchical modulation recognition system was introduced in \cite{channelfading} which shows that with increased path fading, the classification accuracy degrades. Some other efforts have utilized machine learning techniques such as support vector machines (SVM) \cite{SVM2}. Other techniques include feature engineering methods obtained using cyclostationarity \cite{cyclo} and wavelet transform \cite{wavelet}. Extracting a proper set of features for classification also has many practical issues. For example, without prior knowledge, the instantaneous phase or frequency cannot be estimated. Work in \cite{cn_cnn} utilizes different variants of CNN architectures to improve the modulation recognition accuracy. A detailed survey on the methods for modulation recognition is presented in \cite{survey}. For fair comparison, we evaluate our scheme on the same dataset that has been used in the prior work \cite{survey}. Correction of channel artifacts in the signal was not considered in prior works due to which even for SNRs greater than 0dB the accuracy reported was poor. The proposed CM+CNN framework addresses this issue by using a learnable correction module in tandem with a CNN, leading to a higher accuracy.

\begin{figure}
\centering
\includegraphics[width = 3.2in]{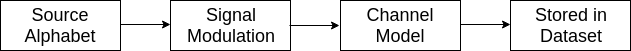} 
\caption{\label{data_generation} \small Data Generation Scheme \cite{dataset}.} 
\end{figure}

For the purpose of developing machine learning models for radio recently an open source, synthetically generated dataset (RadioML2016.10a) using GNUradio was introduced \cite{dataset}. Fig.\ref{data_generation} illustrates the dataset generation technique. The channel incorporates a sampling frequency offset, carrier frequency offset and a phase noise using a random walk process. Additive white gaussian noise (AWGN) further degrades the signal. Parameters used to model the channel are listed in Table \ref{table_chan}. The model for signal generation is a complex enough replication of real radio transmission signals making it a quality dataset for developing algorithms and performing simulations for software based radio. 

\begin{table}
\begin{center}
\begin{tabular}{|l|c|r}
\hline
\textbf{Parameter} & \textbf{Value} \\
\hline
Sampling frequency & 200 kHz\\
\hline
Sampling rate offset standard deviation  & 0.01 Hz\\
\hline
Maximum sampling rate offset& 50 Hz\\
\hline
Carrier frequency offset standard deviation & 0.01 Hz\\
\hline
Maximum carrier frequency offset & 500 Hz\\
\hline
Number of sinusoids used in frequency selective fading  & 8\\
\hline
Maximum doppler frequency used in fading & 1\\
\hline
Fading model & Rician\\
\hline
Rician K-factor & 4\\
\hline
Delays & [0.0, 0.9, 1.7]\\
\hline
Magnitudes corresponding to each delay time & [1, 0.8, 0.3]\\
\hline
Ntaps & 8\\
\hline
Standard deviation of the AWGN process & $10^{-\frac{\text{SNR}}{10} }$\\
\hline

\end{tabular}
\caption{Channel Model Parameters \cite{dataset}}
\label{table_chan}
\end{center}
\end{table}


\section{Proposed Technique}
\label{technique}
\begin{figure*}
\centering
\includegraphics[width=6in]{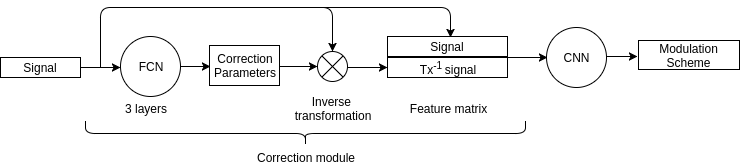} 
\caption{\label{network} \small Correction module and network architecture.}  
\end{figure*}
As described in Section \ref{intro}, addition of CFO and phase noise to the signal hampers modulation recognition. In this section we introduce a correction module (CM) to address this issue as depicted in Fig. \ref{network}. Overall the CM can be divided into two parts. The first part is a trainable function that estimates the phase and frequency offsets (correction parameters) from the received signal. The second part is a static function that generates the input for CNN by undoing the frequency and phase distortion on the received signal using the offsets estimated by the first part. The first part is trained by backpropagating the error from the modulation recognition label through the CNN and through the second part. Thus, no additional supervised information is needed such as true phase or frequency offset. 

\subsection{Correction parameter estimation}\label{aa}
For the first part of the CM, we utilize a fully connected network (FCN) to estimate CFO $\omega$ and phase offset $\phi$. The FCN has one hidden layer (in which 80 hidden neurons gave good validation performance) followed by a final layer with two outputs $\omega$ and $\phi$. To allow the estimation of these two parameters from a continuous and unbounded range assuming no prior knowledge, we choose the activation function of the final layer to be linear. Since the signal is distorted randomly, the error in the estimation of the correction parameters may vary with SNR and modulation scheme. Therefore we also experimented with the idea of simultaneously giving multiple versions of the signal to the CNN along with the original (uncorrected) signal as well. Assuming that there are $K+1$ pairs of correction parameters $(\omega_{k},\phi_{k})$ indexed by $k$, such that $k=0$ was reserved for the received signal without any estimated correction (i.e. $(\omega_{0},\phi_{0})=(0,0)$, and rest of the $K$ signals were generated using the $2K$ output neurons of the FCN.

\subsection{Generating input for CNN}\label{ab}
The second part of the module applies the phase and frequency inverse transformations using the estimated correction factors by multiplying the received signal $x_n$ with $e^{-j\omega_{k}n - j \phi_{k}}$, where $n$ is the discrete time index of the signal. That is, the second part implemented the following equations:
\begin{align}
    Y_{k,n}^{(I)} = \Re\left( x_n e^{-j\omega_{k}n - \phi_{k}} \right) \label{def_ykn1}\\
    Y_{k,n}^{(Q)} = \Im\left( x_n e^{-j\omega_{k}n - \phi_{k}} \right) \label{def_ykn2}
\end{align}
In practice, we obtained best results with $K=1$ as shown in Fig. \ref{network}. That is, $k=0$ corresponded to the original signal, while we needed to estimate only one frequency-phase pair for $k=1$ using the FCN part requiring it to have only two output neurons. Thus, the dimension of the output of this part was $128\times 2(K+1)$, where the dataset had $128$ samples for each signal, and the factor $2$ accounts for both real and imaginary parts of the signal. Thus, the output of the CM was sized $128 \times 4$.

\subsection{End-to-end training and CNN architectures}
The output of the CM, which was $K+1$ versions of the received signal, was input into the CNN that estimated the modulation scheme. To train CM+CNN, i.e. the parameters of the FCN and the CNN, the recognition error was backpropagated through the cascade of CNN, inverse transformation, and FCN. This was possible because the sub-gradient of outputs of the FCN and the CNN with respect to their respective inputs and parameters (weights and biases) exists everywhere by design. Additionally, a quick look at \eqref{def_ykn1}, \eqref{def_ykn2} is sufficient to realize that the gradient of the outputs of the inverse transformation with respect to its inputs $x_n$, $\omega_k$, and $\phi_k$ also exists. This allowed end-to-end backpropagation using only the knowledge of the modulation scheme for the training data without additional knowledge of the actual frequency or phase offsets thus learning it unsupervised. 

To improve the modulation recognition accuracy, we trained two different CNNs, one each for SNR below and above $0$ dB. Based on prior studies we assume that whether the SNR is above or below 0dB can be determined, even without the knowledge of the modulation scheme \cite{SNR1},\cite{SNR2}. We confirmed that architectures similar to the ones described previously in \cite{survey} worked well, which is satisfying as it also allows direct assessment of adding the proposed correction module (CM). We used a four-layer CNN for non-negative SNR and a three-layer CNN for negative SNR based on a validation process. All convolutional layers of our CNNs (whether three or four) had $50$ one-dimensional convolutional filters of size $c_{l} \times 8$, where $c_{l}$ is the number of feature maps or channels of the previous layer. We used $4$ input channels as described in \ref{ab} for the input layer, unlike the $2$ channels of \cite{survey}. The convolution was performed using \emph{valid} setting and thus no padding was required at signal edges. The first two convolutional were each followed by max-pooling of factor $2$. The output of the convolutional layers is followed by a dense layer having $512$ neurons. The output layer had $11$ neurons. All layers used rectified linear activation, except the output layer that used softmax.

\begin{figure}
\centering
\includegraphics[width=3.2in]{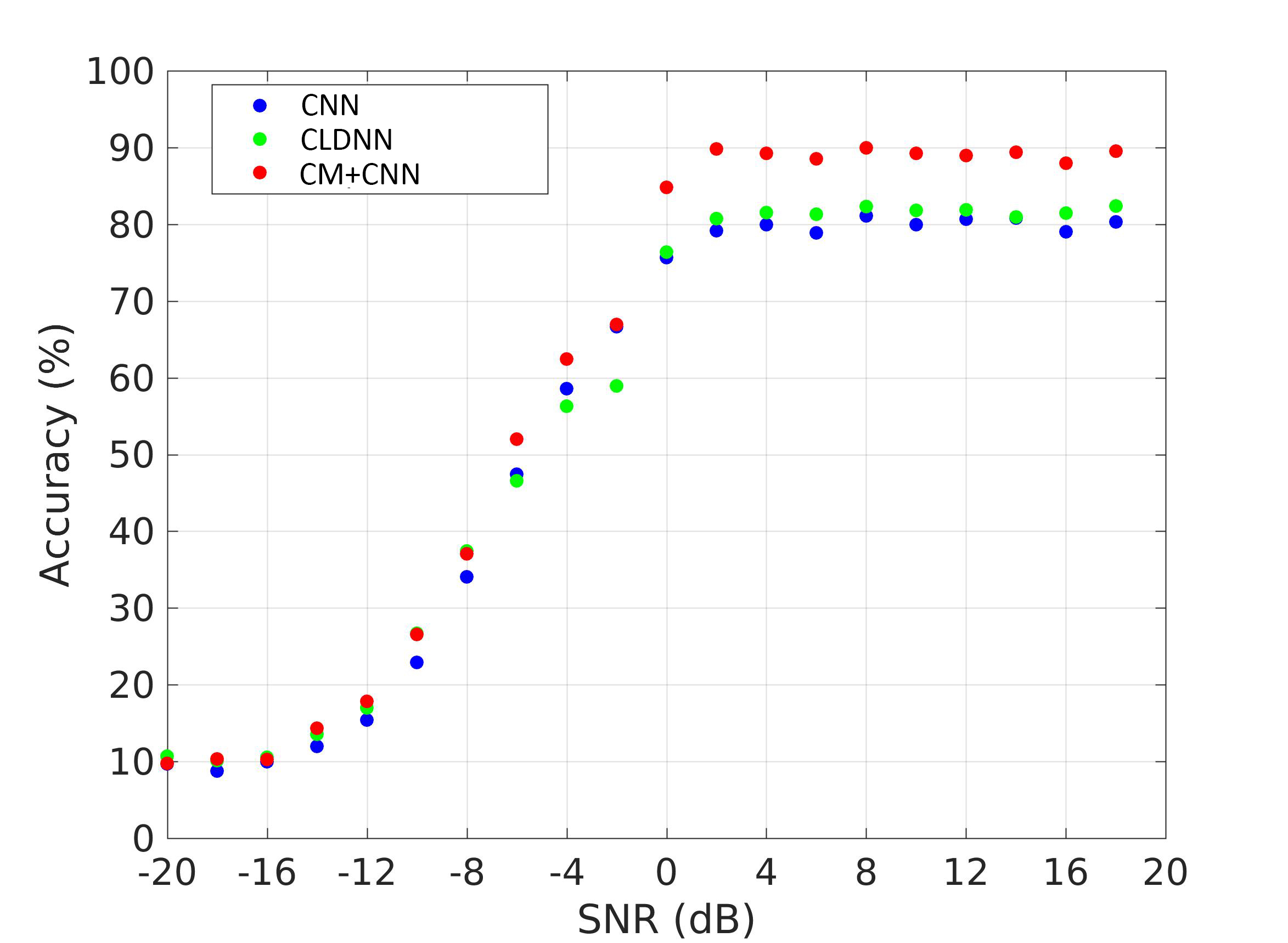} 
\caption{\label{compare} \small Accuracy comparison between proposed technique (CM+CNN) and previous benchmark (CNN, CLDNN)}  
\end{figure}
\begin{figure}
\centering
\hspace{-0.55cm}
\includegraphics[width=3.43in]{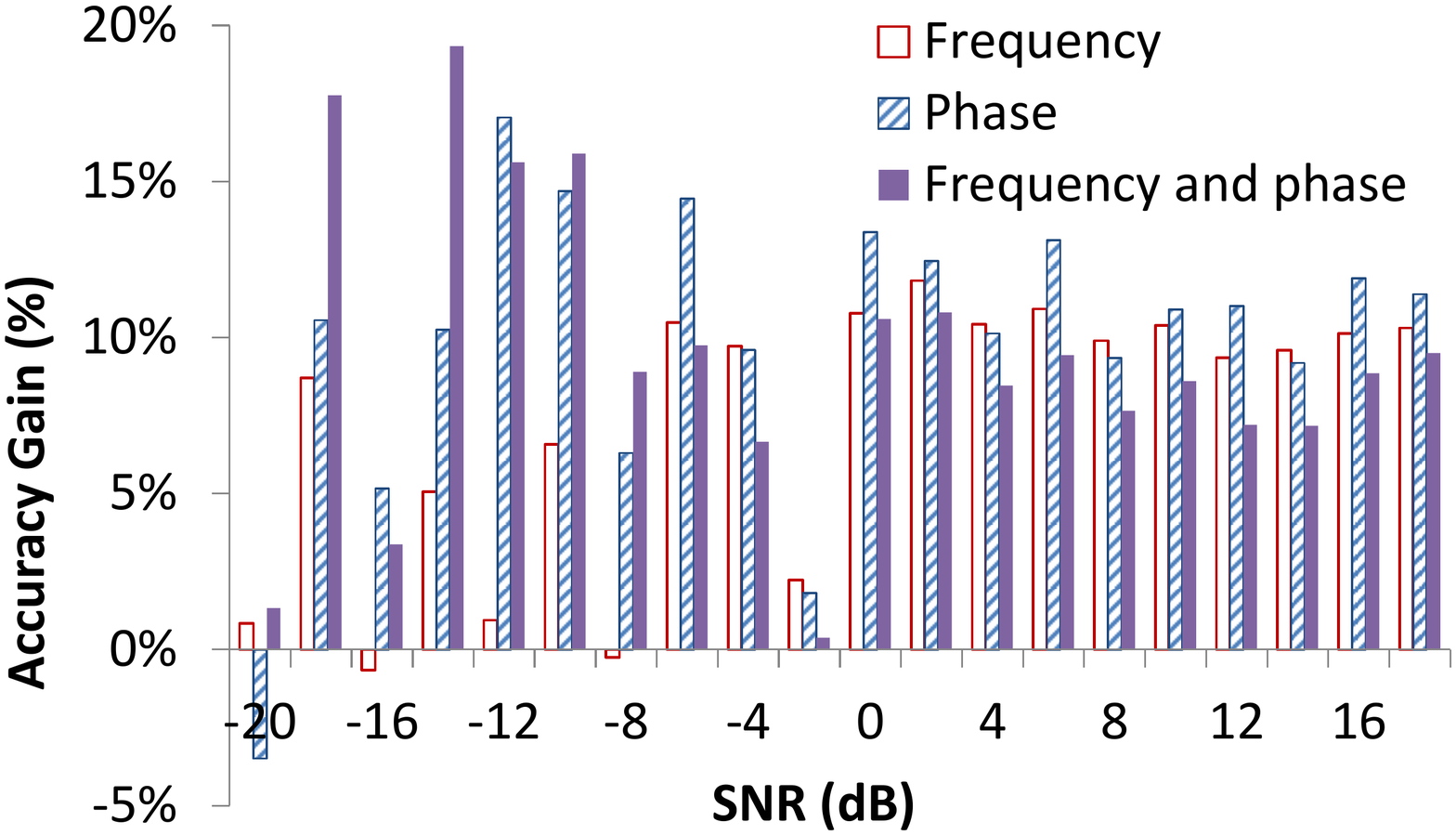} 
\caption{\label{gain1} \small Accuracy gain with respect to the base CNN for frequency correction, phase correction, both frequency and phase corrections}  

\end{figure}
\section{Results and Discussion}
\label{results}
Among various experiments we conducted to determine the useful combinations of neural network architectures and hyperparameters, including the number of estimated correction parameters, we describe those that led to conclusive results. The RadioML2016.10a dataset that we used has signals with 11 analog and digital modulation schemes with SNR varying from $-20$dB to $+18$dB. Since every signal passes through the channel described in Table \ref{table_chan}, it gets distorted by sampling rate offset, carrier frequency offset, phase noise and AWGN. The correction module in our work accounts for phase and frequency offset. We also verified the previous benchmark results by reimplementing CNN and CLDNN \cite{survey}. The architecture used for CNN has 3 convolutional layers each having $50$ filters with $1 \times 8$ filter size. For CLDNN the output of $3$ convolutional layers is concatenated with the output of the first convolutional layer. Comparison of modulation recognition accuracy between the proposed method, CNN and CLDNN for different SNRs is shown in Fig. \ref{compare}. For SNRs above -$14$dB a higher accuracy is observed using the proposed technique with significant improvements for SNR greater than 0dB. Similar performance improvement is observed for SNR less than -$14$dB. 

We also experimented with the following three cases of parameter corrections: 1) frequency only, 2) phase only, and 3) frequency and phase corrections. Accuracy gains with respect to the base CNN for the three cases are presented in Fig. \ref{gain1}. We observed significant gains for nearly all the cases, thus demonstrating the benefit of frequency and phase offset corrections. 

%
%

\begin{figure}
\centering
\includegraphics[width=2.9in]{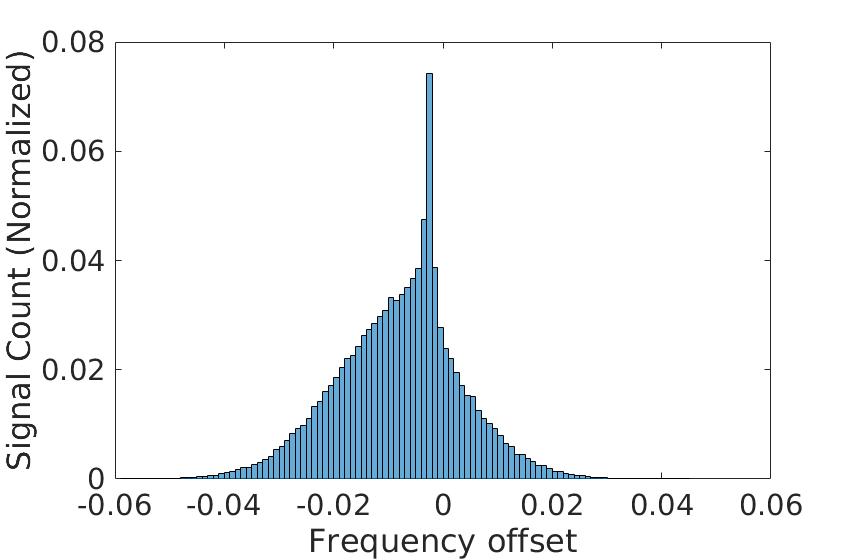} 
\caption{\label{om} \small Normalized histogram for frequency corrections}  
\end{figure}

\begin{figure}[t!]
   \begin{subfigure}{0.5\textwidth}
   \centering
    \includegraphics[width=2.8in]{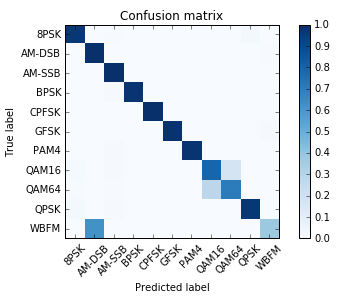} 
    \caption{}
    \label{cnfp}
    \end{subfigure}
     ~
    \begin{subfigure}{0.5\textwidth}
    \centering
    \includegraphics[width=2.8in]{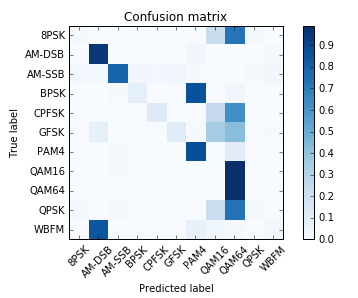} 
    \caption{}  
    \label{cnfn}
    \end{subfigure}
\caption{\label{cnf} (a) Confusion matrix for non-negative SNR, (b) Confusion matrix for negative SNR}
\end{figure}

Fig. \ref{om} shows the output of the correction module. Note that the activation function of the final fully connected layer of the FCN was linear as described in Section \ref{aa}. The output could have been any real value. But we observe in Fig. \ref{om} most of the frequency corrections lie in the range of $-0.01$ Hz to $+0.01$ Hz. The standard deviation of the frequency corrections obtained is $0.01131$ Hz. This matches with the standard deviation of carrier frequency offset that is used to model the channel as listed in Table \ref{table_chan}. Hence the correction module estimates the CFO closely with the actual offset values without any extra supervised data. Due to complex selective channel fading and delays, it is difficult to estimate the actual range of random phase noise. In our experiments we found the phase noise correction to vary between $150$$^{\circ}$  and $270$$^{\circ}$ with mode at $240$$^{\circ}$. Further we plot the confusion matrix for non-negative and negative SNRs in Fig. \ref{cnf}. For non-negative SNRs we observe that the major confusion is between QAM16 and QAM64. A reason for this can be that features for a signal with QAM64 modulation may not be captured by just 128 samples due to which the deep network confuses it with QAM16. Due to increased noise we observed confusion to have increased for negative SNR signals. All the techniques do no better than a random guess for signals having SNR lower than $-14$ dB as shown in Fig. \ref{compare}.  


\section{Conclusion}
\label{concl}
We introduced a new module in this paper to estimate the carrier frequency offset and phase noise of the received signal to improve modulation recognition accuracy. The proposed network outperforms the previous benchmark achieving significant accuracy improvements for both high and low SNR signals. Since a generic CNN is not designed to deal with the effects caused by wireless channels, we addressed this issue by introducing a correction module. Further we observe that the frequency corrections calculated corresponds closely with the actual frequency offsets caused by the channel. Since there can be any number of perceptrons (correction factors) in the final layer, corrections other than phase or frequency can also be estimated. We have demonstrated that the concept of spatial transformer networks \cite{stn} can be generalized to distortion correction for signals in a cognitive radio setup. Similarly, distortion parameters for audio and speech signals can also be estimated for signal correction before a recognition task in a similarly cascaded neural network that can be trained end-to-end.  

\bibliographystyle{IEEEbib}
\bibliography{citations}

\begin{thebibliography}{10}

\bibitem{survey}
N.~West and T.~O'Shea,
\newblock ``Deep architectures for modulation recognition,''
\newblock {\em IEEE International Conference on Dynamic Spectrum Access
  Networks}, 2017.

\bibitem{spectrum_policy}
S.~Haykin,
\newblock ``Cognitive radio: brain-empowered wireless communications,''
\newblock {\em IEEE Journal on Selected Areas in Communications}, vol. 23, pp.
  201--220, 2005.

\bibitem{spectrum_sensing}
T.~Yucek and H.~Arslan,
\newblock ``A survey of spectrum sensing algorithms for cognitive radio
  applications,''
\newblock {\em IEEE Communications Surveys \& Tutorials}, vol. 11, pp.
  116--130, 2009.

\bibitem{timede}
O.B. Akan, O.B. Karli, and O.~Ergul,
\newblock ``Cognitive radio sensor networks,''
\newblock {\em IEEE Network}, vol. 23, 2009.

\bibitem{DL}
Y.~LeCun, G.~Bengio, and G.~Hinton,
\newblock ``Deep learning,''
\newblock {\em Nature}, vol. 521, 2015.

\bibitem{nandi}
AK~Nandi and EE~Azzouz,
\newblock ``Algorithms for automatic modulation recognition of communication
  signals,''
\newblock {\em IEEE Transactions on communications}, vol. 46, 1998.

\bibitem{stn}
M.~Jaderberg, K.Simonyan, et~al.,
\newblock ``Spatial transformer networks,''
\newblock {\em Advances in Neural Information Processing Systems}, 2015.

\bibitem{channelfading}
E.~Like and V.~Chakravarthy,
\newblock ``Signal classification in fading channels using cyclic spectral
  analysis,''
\newblock {\em Journal on Wireless Communications and Networking}, 2009.

\bibitem{SVM2}
S.~Hassanpour, A.~Pezeshk, and F.~Behnia,
\newblock ``Automatic digital modulation recognition based on novel features
  and support vector machine,''
\newblock {\em International Conference on Signal-Image Technology \&
  Internet-Based Systems}, 2016.

\bibitem{cyclo}
O.~Dobre, A.~Abdi, et~al.,
\newblock ``Survey of automatic modulation classification techniques: classical
  approaches and new trends,''
\newblock {\em Communications, IET}, vol. 1, pp. 137--156, 2007.

\bibitem{wavelet}
P.~Prakasam and M.~Madheswaran,
\newblock ``Digital modulation identification model using wavelet transform and
  statistical parameters,''
\newblock {\em Comp. Sys., Netw., and Comm}, 2008.

\bibitem{cn_cnn}
T.~O’Shea, J.~Corgan, and T.~Clancy,
\newblock ``Convolutional radio modulation recognition networks,''
\newblock {\em Engineering Applications of Neural Network}, 2016.

\bibitem{dataset}
T.~O'Shea and N.~West,
\newblock ``Radio machine learning dataset generation with gnu radio,''
\newblock {\em Proceedings of the GNU Radio Conference}, 2016.

\bibitem{SNR1}
D.~Wu, X.~Gu, and Q.~Guo,
\newblock ``Blind signal-to-noise ratio estimation algorithm with small samples
  for wireless digital communications,''
\newblock {\em Intelligent Computing in Signal Processing and Pattern
  Recognition, Springer, Berlin, Heidelberg}, 2006.

\bibitem{SNR2}
S.~Dan and G.~Lindong,
\newblock ``A blind snr estimator for digital bandpass signals,''
\newblock {\em Journal of Electronics}, vol. 25, 2008.

\end{thebibliography}
\end{document}